\documentclass[twocolumn,showpacs,noshowkeys,floatfix,prd,aps]{revtex4}
\usepackage{epsfig} 
\usepackage{longtable} 
\usepackage{hyperref}
\begin{document}  
 
\begin{titlepage} 
\null 
\vspace{2cm} 
\begin{center} 
\Large\bf  
The hadronic $\tau$ decay of a heavy charged Higgs \\ 
in models with singlet neutrino in large extra dimensions 
\end{center} 
\vspace{1.5cm} 
 
\begin{center} 
\begin{large} 
K\'et\'evi~Adikl\`e~Assamagan\\ 
\end{large} 
\vspace{0.5cm} 
Department of Physics, Brookhaven National Laboratory\\ 
Upton, NY 11973 USA\\ 
\vspace{0.7cm} 
\begin{large} 
Aldo Deandrea\\ 
\end{large} 
\vspace{0.5cm} 
Institut de Physique Nucl\'eaire, Universit\'e de Lyon I\\  
4 rue E.~Fermi, F-69622 Villeurbanne Cedex, France 
\end{center} 
 
\vspace{1.3cm} 
 
\begin{center} 
\begin{large} 
{\bf Abstract}\\[0.5cm] 
\end{large} 
\parbox{14cm}{We study the LHC sensitivity to the charged Higgs discovery in  
the channel $H^-\rightarrow\tau_L^-\nu$ in models with a singlet neutrino in 
large extra dimensions. The observation of such a signal would provide a 
distinctive evidence for these models since in the standard two Higgs doublet 
model type II, $H^-\rightarrow\tau_L^-\nu$ is completely suppressed.} 
\end{center} 
 
\vspace{2cm} 
\vfil 
\noindent 
LYCEN-2001-77\\ 
November 2001\\ 
\vfill 
\eject 
\end{titlepage} 
 
\newpage 
 
\title{The hadronic $\tau$ decay of a heavy charged Higgs \\ 
in models with singlet neutrino in large extra dimensions}  
\author{K\'et\'evi~A. Assamagan}  
\email{ketevi@bnl.gov} 
\affiliation{Department of Physics, Brookhaven National Laboratory, 
Upton, NY 11973 USA}
\author{Aldo Deandrea}
\email{deandrea@ipnl.in2p3.fr}
\affiliation{Institut de Physique Nucl\'eaire, Universit\'e Lyon I, 4 rue
E.~Fermi,  F-69622 Villeurbanne Cedex, France}   
\date{November, 2001}  
\preprint{LYCEN-2001-77} 
\begin{abstract}  
We study the LHC sensitivity to the charged Higgs discovery in the channel  
$H^-\rightarrow\tau_L^-\nu$ in models with a singlet neutrino in 
large extra dimensions. The observation of such a signal would provide a 
distinctive evidence for these models since in the standard two Higgs doublet 
model type II, $H^-\rightarrow\tau_L^-\nu$ is completely suppressed.  
\end{abstract}  
 
\maketitle  
 
\section{Introduction}  
The possibility that our world has more than four space--time dimensions 
has been considered long time ago \cite{KK}. More 
recently phenomenological studies based on simplified models have brought new 
insight on how extra dimensions may show up in present and future experimental 
setups. Localization of Standard Model (SM) degrees of freedom on a 
(3+1)--dimensional wall or 3--brane explains why low energy physics is 
effectively four dimensional \cite{1}. In models where extra dimensions open up 
at the TeV scale, small neutrino masses can be generated without implementing 
the seesaw mechanism \cite{2}. These models postulate the existence of 
$\delta$ additional spatial dimensions of size $R$ where gravity and perhaps 
other fields freely propagate while the SM degrees of freedom are confined to 
(3+1)-dimensional wall (4D) of the higher dimensional space. The idea that our
world could be a topological defect of a higher--dimensional theory~\cite{rub}
finds a natural environment in string theory~\cite{pol}.

The true scale of gravity, or fundamental Planck scale $M_*$, of the ($4+\delta$)D 
space time is related to the reduced 4D Planck scale $M_{Pl}$, as:   
\begin{equation} 
\label{eq:scale}  
M_{Pl}^2 = R^\delta M_*^{\delta+2}\; ,
\end{equation}  
where $M_{Pl}=2.4 \times 10^{18}$ GeV is related to the usual Planck mass
$1.2 \times 10^{19}$ GeV  $=\sqrt{8\pi} M_{Pl}$.
Since no experimental deviations from Newtonian gravity are 
observed at distances above 0.2 mm \cite{expgra}, the extra dimensions must be 
at the sub-millimeter level with $M_*$ as low as few TeV and $\delta \geq 2$.  
 
The right handed neutrino can be interpreted as a singlet with no quantum 
numbers to constrain it to the SM brane and thus, it can propagate into the 
extra dimensions just like gravity~\cite{3}. Such singlet states in the bulk 
couple to the SM states on the brane as right handed neutrinos with small 
couplings -- the Yukawa couplings of the bulk fields are suppressed by the 
volume of the extra dimensions. The interactions between the bulk neutrino  
and the wall fields generate Dirac mass terms between the wall fields and all 
the Kaluza-Klein modes of the bulk neutrino. As long as this mass is less than 
$1/R$, the Kaluza-Klein modes are unaffected while for the zero mode, the 
interaction generates a Dirac neutrino mass suppressed by the size of the 
extra dimensions:  
\begin{equation}  
\label{eq:dm}  
m_D =\frac{\lambda}{\sqrt{2}}\frac{M_*}{M_{Pl}}v  
\end{equation}  
where $\lambda$ is 
a dimensionless constant and $v$ the Higgs vacuum expectation value (VEV),
$v\simeq  246$~GeV. The mixing between the lightest neutrino with mass $m_D$
and the  heavier neutrinos introduces a correction $N$ to the Dirac mass such
that the  physical neutrino mass $m_\nu$ is~\cite{2}:  
\begin{equation}   
\label{eq:nu}  
m_\nu =\frac{m_D}{N},   
\end{equation}   
where  
\begin{equation}  
\label{eq:N}  
N \simeq 1 + \sum_{\vec{n}}^{|\vec{n}|< M_* R}  \left(\frac{m_D
R}{\vec{n}}\right)^2 \; ,   
\end{equation}  
$\vec{n}$ is a vector with $\delta$ integer components counting the number of
states and the summation is taken over the Kaluza-Klein states up the
fundamental scale $M_*$.  The sum over the different Kaluza-Klein states can be
approximately replaced by a continuous integration. The following formula can
be used: 
\begin{equation}
\label{intsum}
\sum_{\vec{n}}^{|\vec{n}|< M_* R} 
f\left(\frac{{\vec{n}}^2}{{R}^2}\right) \longrightarrow
S_\delta R^\delta \int_0^{M_*} dx \,x^{\delta-1} f(x^2)\; ,
\end{equation}
where $f$ is a function of ${\vec{n}}^2/{{R}^2}$ and
$S_\delta=2\pi^{\delta/2}/\Gamma(\delta/2)$ is the surface of a unit radius
sphere in $\delta$ dimensions.  After summing over Kaluza-Klein states up to
the  cut-off $M_*$, assuming $\delta \neq 2$:  
\begin{equation}  
\label{eq:N_prime} N  
\simeq 1 + \left(\frac{m_D}{M_*}\right)^2\left(\frac{M_{Pl}}{M_*}\right)^2 
\frac{2\pi^{\delta/2}}{\Gamma(\delta/2)}\frac{1}{\delta-2} \; .
\end{equation} 
As shown in Table~\ref{tab:table1}, small neutrino masses, $m_\nu$, can be obtained 
consistent with atmospheric neutrino oscillations~\cite{4}.  
\begin{table*} 
\begin{center}  
\begin{minipage}{.92\linewidth} 
\caption{\label{tab:table1}The parameters used in the current analysis of the 
signal with the corresponding polarization asymmetry. In general, $H^-$ would 
decay to $\tau^-_L$ and $\tau^-_R$, $H^-\rightarrow\tau_R^-\bar{\nu} + 
\tau_L^-\psi$, depending on the asymmetry. For the decay 
$H^-\rightarrow\tau^-_R\bar{\nu}$ (as in MSSM), the asymmetry is $-1$ and this case is already studied for the LHC~\cite{6,7}. The 
signal to be studied is $H^-\rightarrow\tau^-_L\psi$.}  
\end{minipage}
\vbox{\offinterlineskip 
\halign{&#& \strut\quad#\hfil\quad\cr    
\colrule 
& &&$M_*$ (TeV) && $\delta_\nu$ 
&& $\delta$ && $m_D$ (eV) &&  $m_{H^\pm}$ (GeV) && $\tan\beta$ && Asymmetry  
&& $m_\nu$ (eV) &\cr 
\colrule  
&Signal-1 && 2 && 4 && 4 && 3.0  &&  219.9 &&  30  && $\sim 1$ && 0.5 $10^{-3}$ 
\cr   
&Signal-2 && 20 && 3 && 3 && 145.0 && 365.4 && 45 && $\sim 1$ && 0.05 &\cr 
&Signal-3 && 1 && 5 && 6 && 5.0 && 506.2 && 4 && $\sim 1$ && 0.05 &\cr   
&Signal-4 && 100 && 6 && 6 && 0.005 && 250.2 && 35&& $\sim -1$ && 0.005 &\cr 
&Signal-5 && 10 && 4 && 5 && 0.1 && 350.0 && 20 && $\sim -1$ && 0.04 &\cr  
&Signal-6 && 50 && 5 && 5 && 0.04 && 450.0 && 25 && $\sim -1$ && 0.04 &\cr  
\colrule
}}   
\end{center}  
\end{table*} 
 
The framework of singlet neutrino in large extra dimensions must satisfy some 
phenomenological constraints: for $\delta=2$, the mixing between the lightest 
state and the higher Kaluza-Klein excitations can be of $O(1)$ and therefore 
problematic since in such a case $m_D <1/R$ is no longer valid. In 
addition, due to such a large mixing, this scenario might run into problem 
with nucleosynthesis~\cite{1,2,3} (we consider $\delta > 2$ in 
this analysis). Finally, too much energy could be dissipated into the bulk 
neutrino modes, leading to an unacceptable expansion rate of the universe if 
$m_D^2 \geq 10^{-3}$~(eV)$^2$ and $1/R \leq 10$~keV~\cite{1,5} (we 
confine this analysis to the parameter space where this constraint is 
satisfied).  
 
The spectrum of many extensions of the SM includes a charged Higgs state. 
We consider as a prototype of these models the 2-Higgs Doublet Model of type 
II (2HDM-II), where the Higgs doublet with hypercharge $-1/2$ couples only to 
right--handed up--type quarks and neutrinos whereas the $+1/2$ doublet couples 
only to right--handed charged leptons and down--type quarks; an example is the 
Minimal Supersymmetric Standard Model (MSSM). In the following we will
continue to use the VEV $v\simeq 246$ GeV as in formula (\ref{eq:dm}). Its
meaning in terms of $v_1$ (VEV of the $+1/2$ doublet) and $v_2$ (VEV of the
$-1/2$ doublet) is the usual one:
\begin{equation}
\frac{v}{\sqrt{2}}=\sqrt{v_1^2+v^2_2}~~~~~~~~ \tan \beta = \frac{v_2}{v_1}
\label{eq:vevs}
\end{equation}

$H^-$ decays to the right handed 
$\tau^-$ through the $\tau$ Yukawa coupling:     
\begin{equation}  
\label{eq:mssm}   
H^-\rightarrow \tau_R^-\bar{\nu}.   
\end{equation}  
The $H^-$ decay to left handed $\tau^-$ is completely suppressed in MSSM. 
However, in the scenario of singlet neutrino in large extra dimensions, $H^-$ 
can decay to both right handed and left handed $\tau^-$ depending on the 
parameters $M_*$, $m_D$, $\delta$, $m_{H^\pm}$ and $\tan\beta$: 
\begin{equation}  
\label{eq:extraD}  
H^- \rightarrow \tau_R^-\bar{\nu} +\tau_L^-\psi,  
\end{equation}  
where $\psi$ is a bulk neutrino and $\nu$ is 
dominantly a light neutrino with a small admixture of the Kaluza-Klein modes 
of the order $mR/|n|$. The measurement of the polarization asymmetry,  
\begin{equation}  
\label{eq:asym}  
A =\frac{\Gamma(H^-\rightarrow\tau_L^-\psi)- 
\Gamma(H^-\rightarrow\tau_R^-\bar{\nu})} 
{\Gamma(H^-\rightarrow\tau_L^-\psi)+ 
\Gamma(H^-\rightarrow\tau_R^-\bar{\nu})}, 
\end{equation}  
can be used to distinguish between the ordinary 2HDM-II and the 
scenario of singlet neutrino in large extra dimensions -- depending on the 
parameters -- since in the 2HDM-II, the polarization asymmetry would be 
$-1.0$. In this framework of large extra dimensions, the polarization 
asymmetry could also be $-1.0$ if the left handed $\tau$ component of the 
decay~(\ref{eq:extraD}) is completely suppressed. In such a case, the decay 
of $H^-$ would be similar to the 2HDM-II case but possibly with a different 
phase space since the neutrino contains some admixture of the Kaluza-Klein 
modes.  
 
The singlet neutrino may not necessarily propagate into the 
$\delta$-extra dimensional space. It is possible to postulate that the singlet 
neutrino propagate into a subset $\delta_\nu$ ($\delta_\nu \leq \delta$) of 
the $\delta$ additional spatial dimensions, in which case the formalism for 
the generation of small Dirac neutrino masses is merely a generalization of 
the case $\delta_\nu=\delta$ discussed above~\cite{1}. 
 
The charged Higgs 
decay to right handed $\tau$, $H^-\rightarrow\tau_R^-\bar{\nu}$ have been 
extensively studied for the LHC~\cite{6,7}. In this paper, we discuss 
the possibility to observe $H^-\rightarrow\tau_L^-\psi$ at the LHC above
the  top-quark mass. Table~\ref{tab:table1} shows the parameters selected for
the  current analysis. The cases where the asymmetry is $+1$ are discussed in 
details. We assume a heavy SUSY spectrum with maximal mixing. The present 
analysis is conducted in the framework of PYTHIA6.1~\cite{8} and 
ATLFAST~\cite{9}, and the Higgs masses and couplings are calculated to 1-loop 
with FeynHiggsFast~\cite{13}.   
 
\section{$H^\pm$ Production and Decays} 
\begin{figure} 
\epsfysize=5truecm 
\begin{center} 
\epsffile{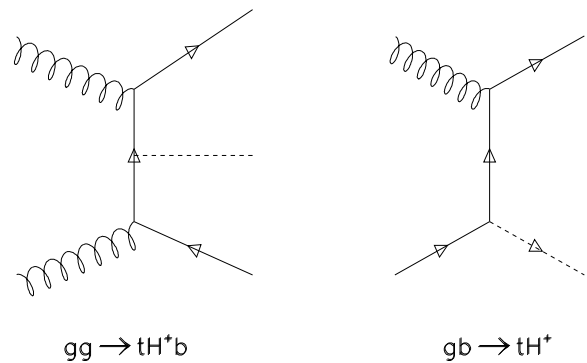} 
\caption{The charged production at the LHC through the $2\rightarrow 3$ process,  
$gg\rightarrow tbH^\pm$ and the $2\rightarrow 2$ process, $gb\rightarrow 
tH^\pm$. The inclusive cross section is the sum of both contributions after 
the subtraction of the common terms. In the framework of large extra 
dimensions with singlet neutrino in the bulk, there are no additional Higgs 
bosons. Thus, the charged Higgs productions are the same as in the 2HDM-II.} 
\label{fig:prod_graphs}  
\end{center}  
\end{figure}

In this framework, no additional Higgs bosons are needed. As a result, the 
charged Higgs productions are the same as in the 2HDM-II, shown in 
Fig.~\ref{fig:prod_graphs}.  
We consider the $2\rightarrow 2$ production 
process where the charged Higgs is produced with a top-quark, $gb\rightarrow 
tH^\pm$. Further, we require the hadronic decay of the top-quark, 
$t\rightarrow Wb\rightarrow jjb$ and the charged Higgs decay to 
$\tau$-leptons. The studies reported in~\cite{6,7} were carried out in 
MSSM where, as previously stated, the $H^-$ would decay to right handed 
$\tau$-leptons: $H^-\rightarrow\tau_R^-\bar{\nu}$. In the scenario of large 
extra dimensions, the $\tau$ decay of charged Higgs would contain both left 
and right handed $\tau$-leptons depending on the asymmetry: 
$H^-\rightarrow\tau_R^-\bar{\nu}+\tau_L^-\psi$. The right handed component, 
$H^-\rightarrow\tau_R^-\bar{\nu}$, is similar to the MSSM case up to some 
phase space factors, in which case the details of the analysis would not 
differ from ~\cite{6,7}. The objective of the current work is to study 
the LHC sensitivity to the left handed component, 
$H^-\rightarrow\tau_L^-\psi$. The detection of such a signal could provide a 
distinctive evidence for models such as large extra dimensions with singlet 
neutrino in the bulk. However, further measurements -- of the rate and the 
polarization asymmetry -- would be necessary to identity the actual scenario 
that is realized.  
 
The major backgrounds are the single top production 
$gb\rightarrow Wt$, and $t\bar{t}$ production with one $W^+\rightarrow jj$ and 
the other $W^-\rightarrow\tau_L^-\bar{\nu}$. Depending on the polarization 
asymmetry (see Equation~\ref{eq:asym}), $H^-\rightarrow\tau_R^-\bar{\nu}$ will 
contribute as an additional background. In Table~\ref{tab:table2}, we list the 
rates for the signal and for the backgrounds.  
\begin{table*} 
\begin{center}  
\begin{minipage}{.78\linewidth} 
\caption{\label{tab:table2}The expected rates ($\sigma\times$ BR), for the signal 
$gb\rightarrow t H^\pm$  with $H^-\rightarrow\tau_R^-\bar{\nu}+\tau_L^-\psi$ 
and $t\rightarrow jjb$, and for the backgrounds:  $W t$ and $t\bar{t}$ 
with $W^-\rightarrow\tau_L^-\bar{\nu}$ and $W^+\rightarrow jj$. We assume an 
inclusive $t\bar{t}$ production cross section of 590~pb. Other cross  sections 
are taken from PYTHIA~6.1 with CTEQ5L parton distribution function.  
See  Table~\protect{\ref{tab:table1}} for the 
parameters used for Signal-1, Signal-2 and Signal-3. In the  last columns, we 
compare the $H^\pm\rightarrow\tau\nu$ branching ratios in this model 
to the corresponding MSSM branching ratios from HDECAY~\cite{10}.} 
\end{minipage} 
\vbox{\offinterlineskip 
\halign{&#& \strut\quad#\hfil\quad\cr 
\colrule 
&Process && $\sigma\,\times\,$~BR (pb) && BR($H^\pm\rightarrow\tau\nu+\tau\psi$) &&  
MSSM: BR($H^\pm\rightarrow\tau\nu$)&\cr  
\colrule  
&Signal-1 && 1.56 && 0.73 &&  0.37  &\cr 
&Signal-2 && 0.15 && 1.0 && 0.15 & \cr  
&Signal-3 && 0.04 && 1.0  &&   0.01 & \cr  
\colrule 
&$t\bar{t}$ && 84.11 &&  &&   &  \cr  
&$gb\rightarrow Wt$ ($p_T>30$~GeV) && 47.56 &&  &&   &  \cr 
\colrule}} 
\end{center}  
\end{table*} 
For the phenomenological analysis, it is convenient to express the partial
widths in terms of inclusive formulas, where the contributions of the
Kaluza-Klein modes are summed up to the kinematical limit $m_\psi \leq m_H$ as
the $\tau$ mass can be neglected.
The partial width of the Higgs decays to
$\tau\nu$ depends of the parameters  $M_*$, $m_D$, $\delta$, $m_{H^\pm}$ and
$\tan\beta$~\cite{agashe}:   
\begin{equation}  
\label{eq:H->t_L}  
\Gamma(H^-\rightarrow\tau_L^-\psi) \simeq 
\frac{m_{H^\pm}}{8\pi}\left(\frac{m_D}{v}\right)^2\frac{\chi_\delta}{\tan^2\beta} 
\left(m_{H^\pm}R\right)^\delta ,  
\end{equation}   
where $(m_{H^\pm}R)^\delta$ is the number of Kaluza-Klein modes lighter than 
the charged Higgs mass and $\chi_\delta$ includes the phase space 
integral:  
\begin{equation} 
\chi_\delta \simeq \frac{2 \pi^{\delta/2}}{\Gamma(\delta/2)}  
\; \left( \frac{1}{\delta}-\frac{2}{\delta+2}+ 
\frac{1}{\delta+4} \right)\; .  
\end{equation} 
Using the relation~(\ref{eq:scale}),  
\begin{equation} 
\label{eq:kk} (m_{H^\pm}R)^\delta = 
\left(\frac{m_{H^\pm}}{M_*}\right)^\delta\times\left(\frac{M_{Pl}}{M_*}\right)^2.  
\end{equation}  
For the $H^-$ decay to the right handed $\tau$, we have~\cite{agashe} 
\begin{eqnarray}  
\label{eq:H->t_R}  
\Gamma&&(H^-\rightarrow\tau_R^-\bar{\nu}) 
\simeq \nonumber \\ 
&&\left[\Gamma(H^-\rightarrow\tau_R^-\bar{\nu})_{\mathrm{MSSM}}\right] 
\frac{\left[1+ f(m_D,M_*,\delta)\right] }{N^2}
\end{eqnarray}  

and the normalization factor $N$ is given by 
Equation~(\ref{eq:N_prime}) and the function $f(m_D,M_*,\delta)$ is (for
$\delta \neq 2$): 
\begin{eqnarray}  
f(m_D,M_*,\delta)&=& 
\frac{m_D^2 \; m_H^{\delta-2}}{M_*^{\delta}}\; 
\left(\frac{M_{Pl}}{M_*}\right)^2\;  
\frac{2 \pi^{\delta/2}}{\Gamma(\delta/2)}  
\nonumber \\  
&\times & \left( \frac{1}{\delta-2}-\frac{2}{\delta}+ 
\frac{1}{\delta+2} \right)\; .  
\label{eq:func} 
\end{eqnarray}  
One can generalize these formulas for a singlet neutrino in a smaller number 
of extra dimensions $\delta_\nu < \delta$ than the extra dimensions available 
to gravity~\cite{2}. Assuming that all the extra dimensions are of the same 
size $R$, one has to replace in formulas (\ref{eq:N_prime})
and (\ref{eq:H->t_L}--\ref{eq:func}):  
\begin{equation} 
\delta \to \delta_\nu~~~~~~~~~~~ \left(\frac{M_{Pl}}{M_*}\right)^2\to 
\left(\frac{M_{Pl}}{M_*}\right)^{2(\delta_\nu/\delta)} 
\end{equation} 
The more general case of a non--symmetric internal $\delta$--dimensional
manifold is given in~\cite{2}.
  
\begin{figure} 
\epsfysize=8truecm 
\begin{center} 
\epsffile{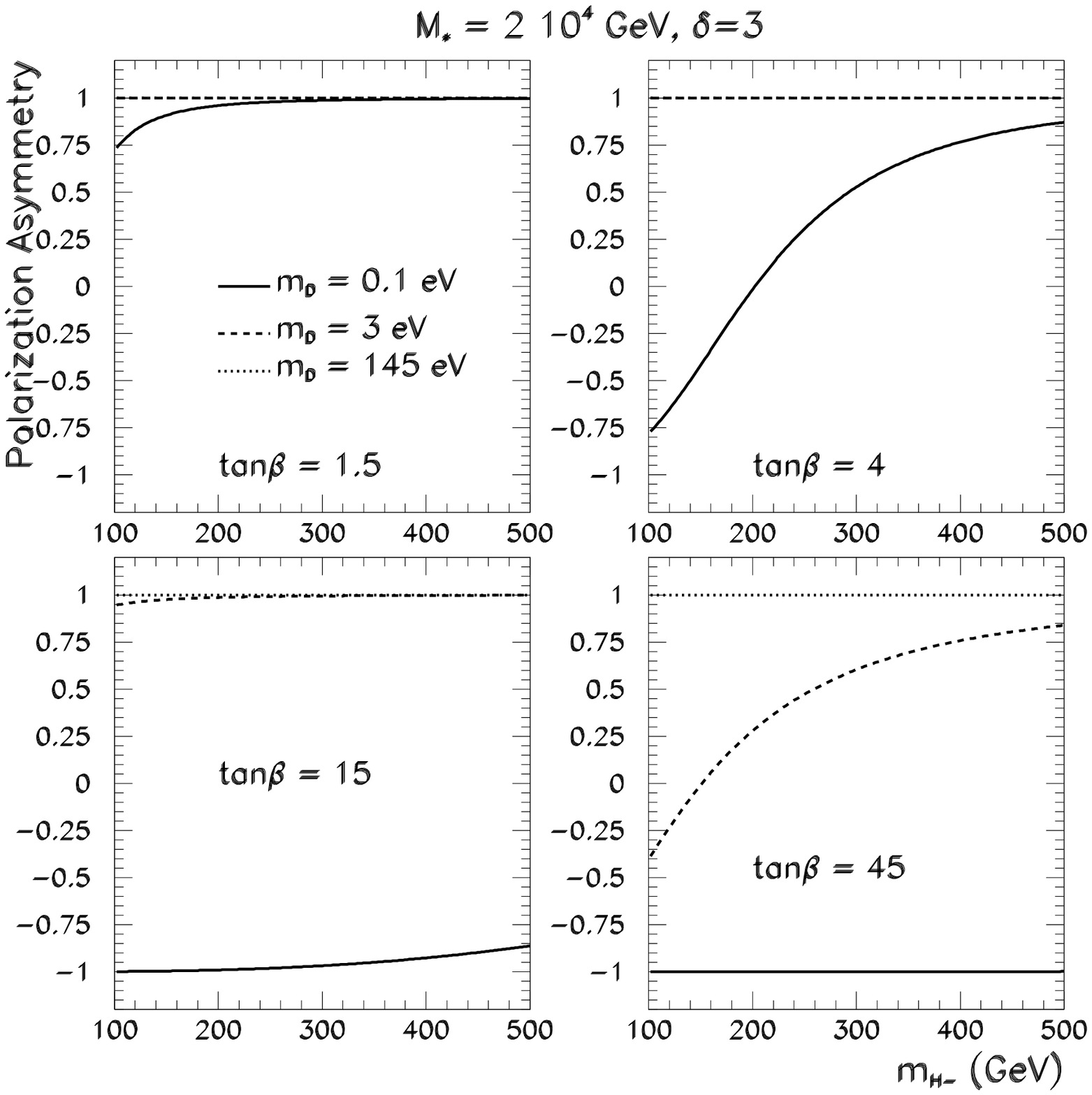} 
\caption{The polarization asymmetry as a function of $m_{H^\pm}$, for various values  
of $\tan\beta$ and $m_D$. For small values of $m_D$, the decay $\tau^-$ are 
right handed (except for small $\tan\beta$ values) while left handed 
$\tau^-$'s are produced as $m_D$ gets larger.}  
\label{fig:tasym_led} 
\end{center}  
\end{figure}

\begin{figure} 
\epsfysize=8truecm 
\begin{center} 
\epsffile{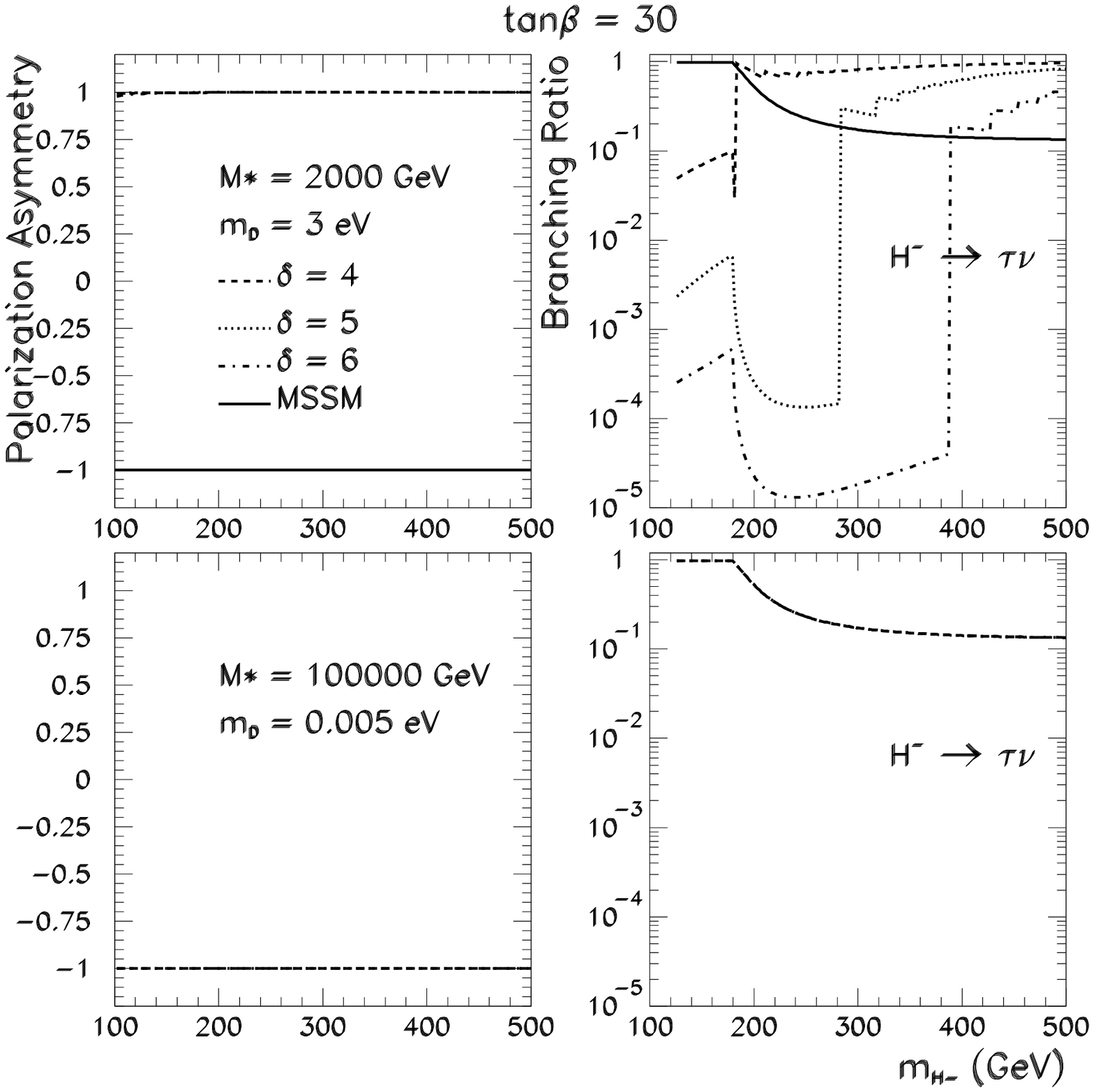} 
\caption{The polarization asymmetry and the $H^\pm\rightarrow\tau\nu$ branching  
ratio for two values of ($M_*$, $m_D$) and $\delta=4$, 5 and 6. For very 
large $M_*$ and small $m_D$, we recovery the MSSM case, i.e., an asymmetry of 
$-1$ (right handed $\tau^-$) and MSSM branching ratios (bottom plots).} 
\label{fig:tasym_led_delta}  
\end{center}  
\end{figure}

Depending on the parameters $M_*$, 
$m_D$, $\delta$, $m_{H^\pm}$ and $\tan\beta$, the $\tau\nu$ decay of the 
charged Higgs can be enhanced or suppressed compared to the MSSM case.  
\begin{figure} 
\epsfysize=8truecm 
\begin{center} 
\epsffile{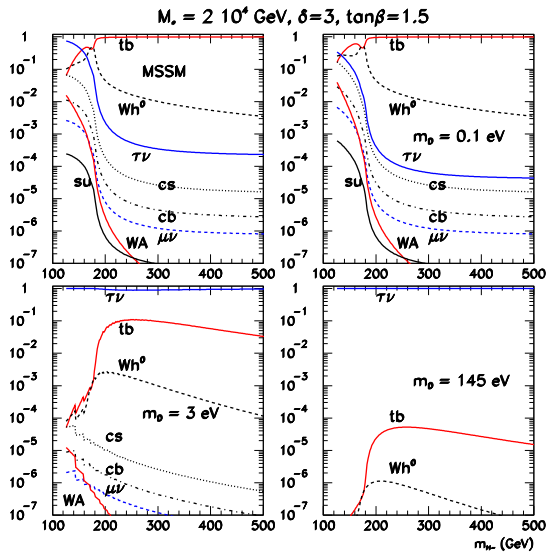} 
\caption{Charged Higgs decays in models with a singlet neutrino in large extra  
dimensions for $M_* = 2\times 10^4$ GeV, $\delta =3$ and $\tan\beta = 1.5$. For 
small values of $m_D$, we see similar decay branchings as in MSSM. As $m_D$ 
gets larger, $H^\pm\rightarrow\tau\nu$ becomes dominant below and above the 
top-quark mass.}  
\label{fig:led_tb1.5}  
\end{center}  
\end{figure} 
\begin{figure} 
\epsfysize=8truecm 
\begin{center} 
\epsffile{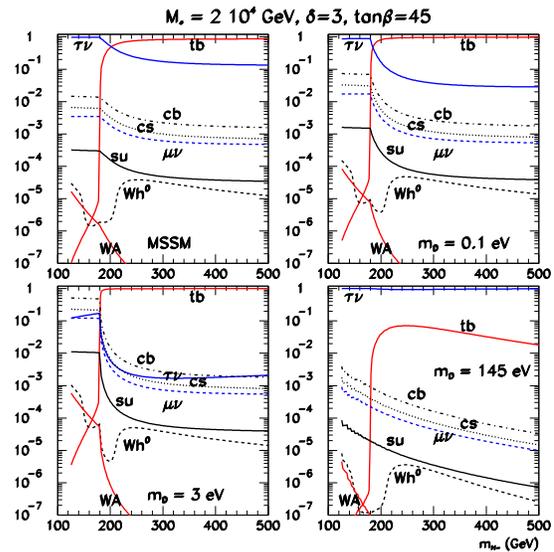} 
\caption{Charged Higgs decays in models with a singlet neutrino in large extra  
dimensions for $M_* = 2\times 10^4$ GeV, $\delta =3$ and $\tan\beta = 45$. The 
dependence in $m_D$ is similar to the situation of 
Fig.~\ref{fig:led_tb1.5}.}  
\label{fig:led_tb45}  
\end{center} 
\end{figure} 
In Fig.~\ref{fig:led_tb1.5} and Fig.~\ref{fig:led_tb45}, we show few 
cases of how the other decays of the charged Higgs are affected in this 
framework; for the chosen values of $M_*$ and $\delta$, the decay branchings 
are similar to MSSM for small values of $m_D$ while at larger $m_D$, the 
$\tau\nu$ decay mode becomes strongly enhanced, especially at low $\tan\beta$. 
In Fig.~\ref{fig:tasym_led},  
we show the polarization asymmetry as a function 
of the charged Higgs mass and for different values of $m_D$ and $\tan\beta$: 
for small $m_D$, right handed $\tau$'s are produced, except at low $\tan\beta$ 
while the asymmetry increases with $m_D$ (see Equation~\ref{eq:H->t_L}). For 
very large values of $M_*$ and small $m_D$, we  recover the MSSM case as shown 
in Fig.~\ref{fig:tasym_led_delta} irrespective of the values of $\delta$ 
considered. 
 
In general, $H^-\rightarrow\tau_L^-\psi+\tau_R^-\bar{\nu}$ 
with the asymmetry between -1 and 1. However, the study of 
$H^-\rightarrow\tau_R^-\bar{\nu}$ has been carried out in detail and reported 
elsewhere~\cite{6,7}. Therefore, in the current study, we consider the 
parameters shown in Table~\ref{tab:table1} and Table~\ref{tab:table2} for 
which the asymmetry is one, i.e., $H^-\rightarrow\tau_L^-\psi$.  
 
\section{Analysis} 
\begin{figure} 
\epsfysize=8truecm 
\begin{center} 
\epsffile{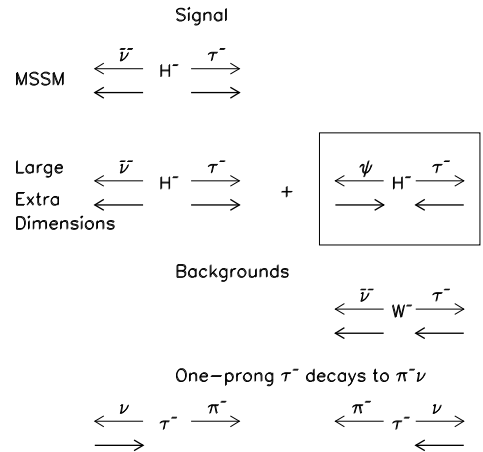} 
\caption{Polarization of the decay $\tau$ from $H^\pm$ in MSSM and in models with  
a singlet neutrino in large extra dimensions. In the latter case, both left 
and right handed $\tau$'s can be produced with some polarization asymmetry. In 
the backgrounds, the $\tau$ comes from the decay of the $W^\pm$. The signal to 
be studied is in the box --- the polarization of the decay $\tau$ in this 
signal is the same as in the background. Thus, $\tau$ polarization effects 
would not help in suppressing the backgrounds but they may help distinguish 
between the 2HDM and other models.}  
\label{fig:tau_pol_led}  
\end{center}  
\end{figure}

\begin{figure} 
\epsfysize=8truecm 
\begin{center} 
\epsffile{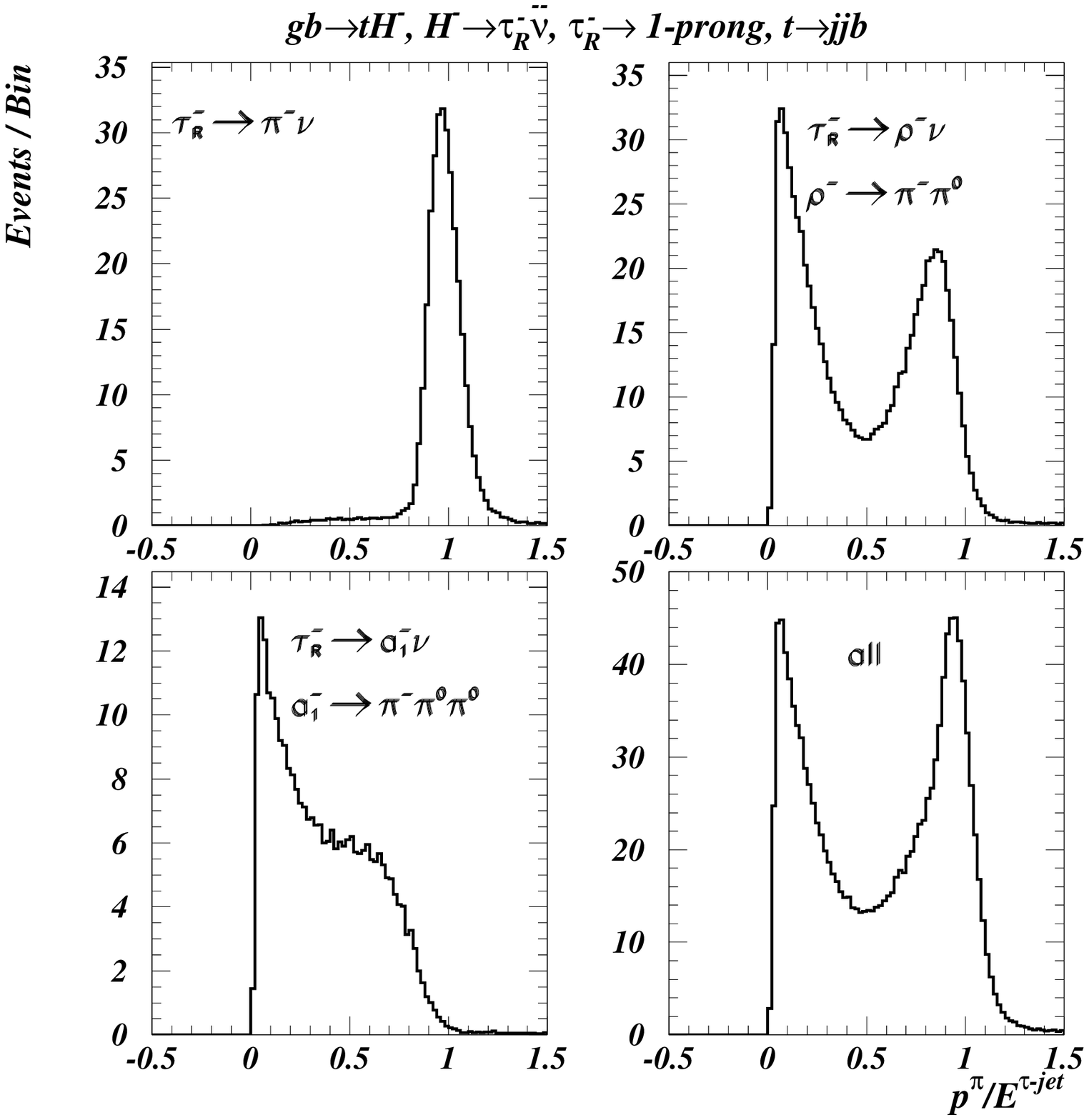} 
\caption{The one prong decays of the $\tau$-lepton from the signal in MSSM:  
$H^-\rightarrow\tau^-_R\bar{\nu}$. We plot the ratio of the momentum carried 
by the charged track to the $\tau$-jet energy. This ratio peaks near 1 for 
$\tau\rightarrow\pi\nu$ and near 0 and 1 for longitudinal $\rho$ and $a_1$. 
For transverse $\rho$ and $a_1$, this ratio peaks in the middle.} 
\label{fig:mssm_1prong}  
\end{center}  
\end{figure}

\begin{figure} 
\epsfysize=8truecm 
\begin{center} 
\epsffile{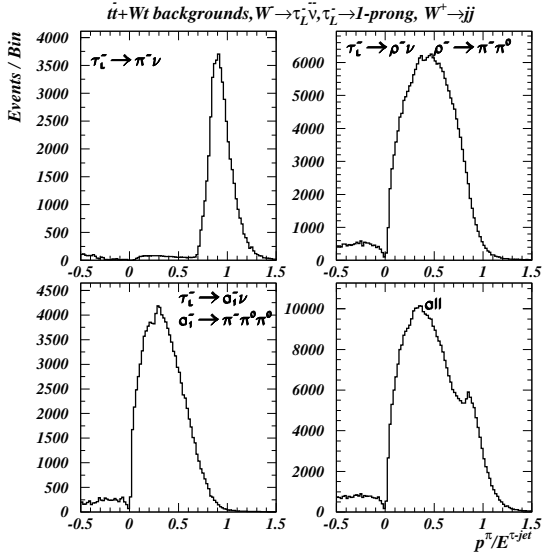} 
\caption{The one prong decays of the $\tau$-lepton from the $t\bar{t}$ and $Wt$  
backgrounds: $W^-\rightarrow\tau^-_R\nu$. Here the situation should be 
reversed and it is for the $\rho$ and $a_1$. For $\tau\rightarrow\pi\nu$, the 
ratio should peak near 0; the peak near 1 comes from the $\tau$-jet labeling 
criteria in ATLFAST: a jet is labeled a $\tau$-jet by requiring the hadronic 
decay products to carry a significant fraction ($>0.9$) of the $\tau$-jet 
energy within a jet cone ($\Delta R < 0.3$). For $\tau\rightarrow\pi\nu$, 
these criteria would select charged pions with this ratio near 1.} 
\label{fig:bgd_1prong}  
\end{center} 
\end{figure}

\begin{figure} 
\epsfysize=8truecm 
\begin{center} 
\epsffile{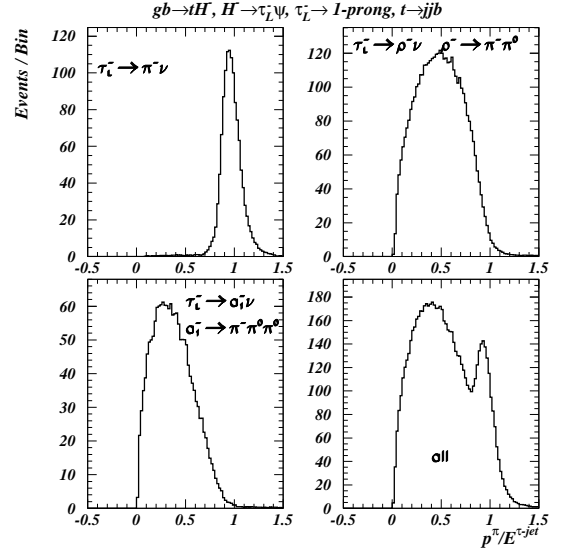} 
\caption{The one prong decays of the $\tau$-lepton from the signal in models with a  
singlet neutrino in large extra dimensions with a polarization asymmetry of 
1: $H^-\rightarrow\tau^-_L\psi$. (The $\tau^-$ from $H^-$ decays are 100\% 
left handed). The situation is thus similar to the backgrounds but opposite to 
signal in MSSM.}  
\label{fig:led_1prong}  
\end{center}  
\end{figure}

The polarization of the $\tau$-lepton is included in this analysis through  
TAUOLA~\cite{11}. We consider the hadronic one-prong decays of the 
$\tau$-lepton since these are believed to carry a better imprint of the 
$\tau$-polarization~\cite{12}:  
\begin{eqnarray}  
\label{eq:pinu} 
\tau^-  \rightarrow & \pi^-\nu  & ~~~(11.1\%) \\ 
\tau^-  \rightarrow & \rho^-(\rightarrow\pi^-\pi^0)\nu  & ~~~(25.2\%) \\ 
\tau^-  \rightarrow & a_1^-(\rightarrow\pi^-\pi^0\pi^0)\nu & ~~~(9.0\%) 
\end{eqnarray} 
In Fig.~\ref{fig:tau_pol_led}, we show the effects of the $\tau$ polarization in the  
signal and the backgrounds in the case of one-prong 
$\tau^-\rightarrow\pi^-\nu$.  
For the signal in MSSM, right handed $\tau_R^-$'s 
come from the charged Higgs decay, $H^-\rightarrow\tau_R^-\bar{\nu}$, while in 
the backgrounds, left handed $\tau_L^-$'s come from the decay of the 
$W^-(\rightarrow\tau_L^-\bar{\nu})$. Since the charged Higgs is a scalar and 
the $W^-$ a vector, the polarization of the $\tau$ results in a stronger 
$\tau$-jet in the MSSM signal than in the backgrounds for 
$\tau^-\rightarrow\pi^-\nu$ and longitudinal $\rho$ and 
$a_1$~\cite{7,12}. The studies reported in ~\cite{6,7} take 
advantage of this polarization effect in suppressing the backgrounds further 
by demanding that the charged track carries a significant part of the 
$\tau$-jet energy:     
\begin{equation}  
\label{eq:sig} 
p_\pi/E^{\tau-\mbox{jet}} > 80\%.  
\end{equation}  
For the signal in MSSM, this 
requirement would retain only the $\pi$ and half of the longitudinal $\rho$ 
and $a_1$ contributions while eliminating the transverse components along with 
the other half of the longitudinal contributions as can be seen from 
Fig.~\ref{fig:mssm_1prong}.  
However, this requirement would suppress much of 
the backgrounds, shown in Fig.~\ref{fig:bgd_1prong}.  
In the framework of 
large extra dimensions, we are interested in $H^-\rightarrow\tau_L^-\psi$ 
where, as shown in Fig.~\ref{fig:tau_pol_led}, the polarization of the 
$\tau$-lepton would be identical to the background case but opposite to the 
MSSM case. Therefore, the requirement~(\ref{eq:sig}) would not help in 
suppressing the backgrounds, as can be seen from Fig.~\ref{fig:bgd_1prong} 
and Fig.~\ref{fig:led_1prong}.  
Nevertheless, there are still some 
differences in the kinematics which can help reduce the background level, and 
we discuss the details of the analysis as follow:  
\begin{description} 
\item[(a)] Search for one-prong hadronic $\tau$ decays with one $\tau$-jet, 
$p_T^{\tau} > 30$~GeV and $|\eta^{\tau}| \leq 2.5$, at least three non $\tau$ 
jets with $p_T^{\mbox{jet}} > 30$~GeV. One of these jets must be a b-tagged 
jet with $|\eta^b| < 2.5$. Further, we apply a b-jet veto by requiring only a 
single b-jet with $|\eta| \leq 2.0$ and $p_T > 50$~GeV. We assume a $\tau$-jet 
identification efficiency of 30\% and a b-tagging efficiency of 50\%, for an integrated luminosity of 100~fb$^{-1}$. We further assume a multi-jet trigger with a high level $\tau$ trigger.
 
\item[(b)] The W from the associated top-quark is reconstructed and the candidates  
satisfying $|m_{jj}-m_W| \leq 25$~GeV are retained (and their four-momenta 
are renormalized to the W mass) for the reconstruction of the top-quark: this 
is done by minimizing the variable $\chi^2 = (m_{jjb}-m_t)^2$. We take $m_W = 
80.14$~GeV and $m_t = 175$~GeV. Subsequently, the events satisfying 
$|m_{jjb}-m_t| < 25$~GeV are retained for further analysis.  
 
\item[(c)] We raise the cut on $p_T^{\tau}$, \textit{i.e.},  
$p_T^{\tau} > 100$~GeV. To satisfy this $p_T^{\tau}$ cut, the $\tau$ jet from  
the backgrounds needs a large $p_T$ boost from the W boson. This will  
result in a smaller opening angle, $\Delta\phi$, between the decay products  
$\tau\nu$. $\Delta\phi$ is the azimuthal opening angle between the $\tau$ jet 
and the missing transverse momentum. In the signal $H^\pm\rightarrow\tau\nu$, 
the $\tau$ jet will require little or no boost at all to satisfy this high 
$p_T^{\tau}$ cut. This explains the backward peak in the $\Delta\phi$ 
distribution  for the signal as shown in Figures~\ref{fig:200_30_misc} 
and~\ref{fig:350_45_misc} -- this backward peak in $\Delta\phi$ is more 
pronounced as the Higgs mass increases.  
\begin{figure} 
\epsfysize=8truecm 
\begin{center} 
\epsffile{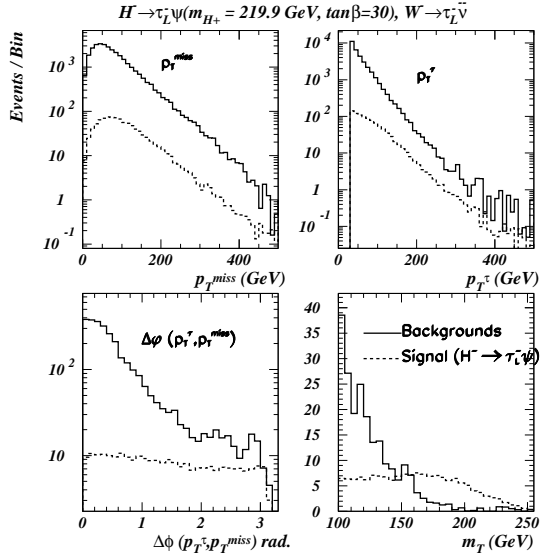} 
\caption{The reconstructions of $p_T^{miss}$, the $\tau$-jet transverse momentum,  
$p_T^{\tau-jet}$ (top plots), the azimuthal opening angle between 
$p_T^{miss}$ and $p_T^{\tau-jet}$ and the transverse charged Higgs mass 
(bottom plots) for the signal, $H^-\rightarrow\tau^-_L\psi$ ($m_A=200$~GeV, 
$\tan\beta=30$) and the backgrounds $W^-\rightarrow\tau^-_L\bar{\nu}$. In the 
signal, the transverse mass is bound from above by the charged Higgs mass 
while in the backgrounds, the transverse mass is constrained by the W mass. 
However, due to the $E_T^{miss}$ resolution, we see a ``leak'' into the signal 
region.}  
\label{fig:200_30_misc}  
\end{center}  
\end{figure} 
\begin{figure} 
\epsfysize=8truecm 
\begin{center} 
\epsffile{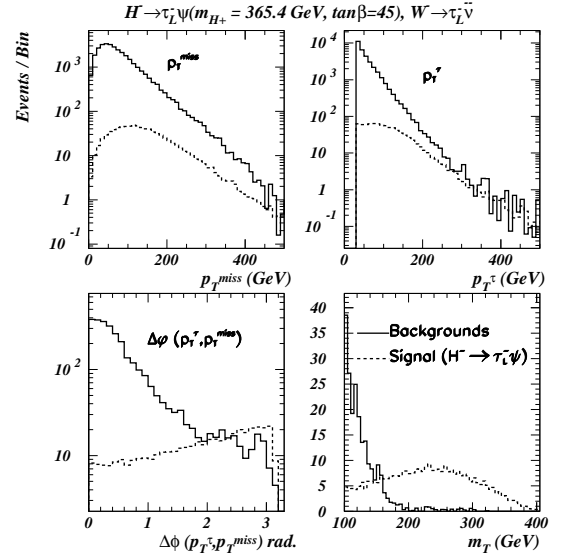} 
\caption{The reconstructions of $p_T^{miss}$, the $\tau$-jet transverse momentum,  
$p_T^{\tau-jet}$ (top plots), the azimuthal opening angle between 
$p_T^{miss}$ and $p_T^{\tau-jet}$ and the transverse charged Higgs mass 
(bottom plots) for the signal, $H^-\rightarrow\tau^-_L\psi$, ($m_A=350$~GeV, 
$\tan\beta=45$) and the backgrounds $W^-\rightarrow\tau^-_L\bar{\nu}$. 
$p_T^{miss}$ and $p_T^{\tau-jet}$ are harder in the signal particularly at 
higher Higgs mass, and the opening angle $\Delta\phi 
(p_T^{\tau-jet},p_T^{miss})$ peaks forward in the backgrounds and backward in 
the signal.}  
\label{fig:350_45_misc}  
\end{center}  
\end{figure} 
Similarly, the missing transverse 
momentum $\not\! p_T$ and the transverse momentum of the $\tau$-jet are 
increasingly harder for the signal as the Higgs mass increases as seen in 
Figures~\ref{fig:200_30_misc} and~\ref{fig:350_45_misc}. Because of the 
neutrino in the final state, only the transverse mass   
\begin{equation} 
\label{eq:trans} m_T = \sqrt{2p_T^\tau\not\! 
p_T\left[1-\cos(\Delta\phi)\right]}  
\end{equation}   
can be reconstructed. In 
the backgrounds, the transverse mass has an upper bound at the $W^-$ mass 
($W^-\rightarrow\tau^-\nu$) while in the signal, it is constrained by the 
charged Higgs mass ($H^-\rightarrow\tau^-\nu$). However, due to the 
experimental resolution  of $E_T^\textrm{\footnotesize miss}$, the $m_T$ 
distribution for the backgrounds shows a ``leak'' into the signal region as 
can be seen in Figures~\ref{fig:200_30_misc} and~\ref{fig:350_45_misc}. 
\item[(d)] To optimize the signal-to-background ratios and the signal 
significances, we apply a cut on the missing transverse momentum: $\not\! p_T 
> 100$~GeV.  
 
\begin{table*} 
\begin{center}  
\begin{minipage}{.55\linewidth} 
\caption{\label{tab:table3}The expected 
signal-to-background ratios and significances calculated after  cut 
\textbf{(e)} for an integrated luminosity of 100~fb$^{-1}$ (one experiment). 
See Table~\ref{tab:table1} for the parameters used for Signal-1, Signal-2 and 
Signal-3. In all the cases considered, the signal can be observed at the LHC 
with significances in excess of 5-$\sigma$ at high luminosity.} 
\end{minipage} 
\vbox{\offinterlineskip 
\halign{&#& \strut\quad#\hfil\quad\cr  
\colrule
& &&Signal-1 && Signal-2 && Signal-3  &\cr 
\colrule
& Signal events  && 41 && 215 && 16 & \cr  
& $t\bar{t}$ && 7 && 7 && 7 & \cr  
& $Wt$ && 3 && 3 && 3 & \cr   
& Total background && 10 && 10 && 10 \cr 
& $S/B$ && 4.1 && 21.5 && 1.6  &\cr  
& $S/\sqrt{B}$ && 13.0 && 68.0 && 5.1 & \cr    
\colrule}}   
\end{center}  
\end{table*}  
\item[(e)] A final cut of $\Delta\phi > 1.0$~rad is applied and the results 
are shown in  Fig.~\ref{fig:led_tnu_bgd} and used for the calculation of the 
signal-to-background ratios and the signal significances shown in 
Table~\ref{tab:table3}.  
\begin{figure} 
\epsfysize=8truecm 
\begin{center} 
\epsffile{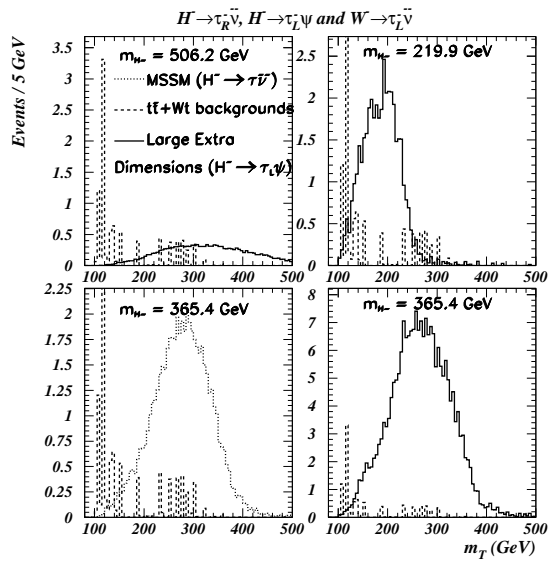} 
\caption{The reconstructions of the transverse mass for the signal in MSSM, the signal  
in models with a singlet neutrino in large extra dimensions and for the 
backgrounds, for an integrated luminosity of 100~fb$^{-1}$. In general, an MSSM charged Higgs can be discovered at the LHC  
depending on $m_A$ and $\tan\beta$. In the 
models with a singlet neutrino in large extra dimensions, the signal can also 
be discovered at the LHC depending on the parameters $M_*$, $\delta$, $m_D$, 
$m_A$ and $\tan\beta$. The observation of the signal in the transverse mass 
distribution would not be sufficient to identify the model: the $\tau$ 
polarization effects must be explored further.}  
\label{fig:led_tnu_bgd} 
\end{center}  
\end{figure} 
The reconstruction of the transverse mass, shown in 
Fig.~\ref{fig:led_tnu_bgd} is not enough to distinguish between the MSSM and 
the singlet neutrino in large extra dimensions. The differences in these two 
scenarios are best seen in the distribution of $p^\pi/E^{\tau-jet}$, the 
fraction of the energy carried by the charged track which is shown in 
Figures~\ref{fig:200_30_0.3ev} and \ref{fig:350_45_145_1}.  
\begin{figure} 
\epsfysize=8truecm 
\begin{center} 
\epsffile{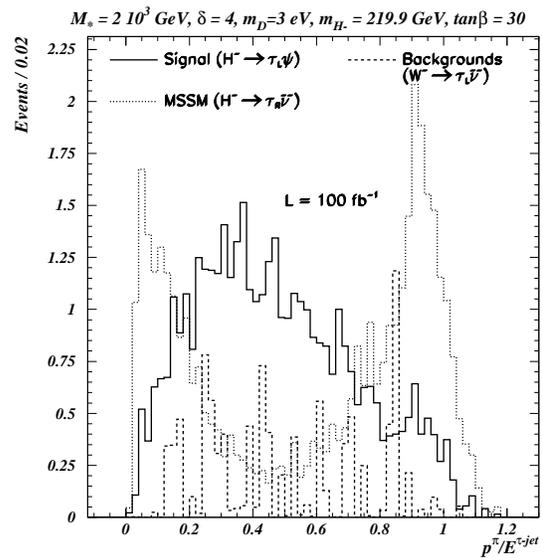} 
\caption{The distribution of the ratio of the charged pion track momentum in one  
prong $\tau$ decay to the $\tau$-jet energy for $m_A=200$~GeV, 
$\tan\beta=30$, $M_*= 2$~TeV, $\delta=4$ and $m_\nu=0.5\,10^{-3}$~eV. In the 
case shown here, the polarization asymmetry is $\sim 1$ ($\sim$ 100\% left 
handed $\tau^-$). We see the difference between MSSM and large extra 
dimensions with singlet neutrino in the bulk. In 
Fig.~\ref{fig:350_45_145_1}, the difference between these two models is more 
pronounced due to the more significant signals.}  
\label{fig:200_30_0.3ev} 
\end{center}  
\end{figure} 
\begin{figure} 
\epsfysize=8truecm 
\begin{center} 
\epsffile{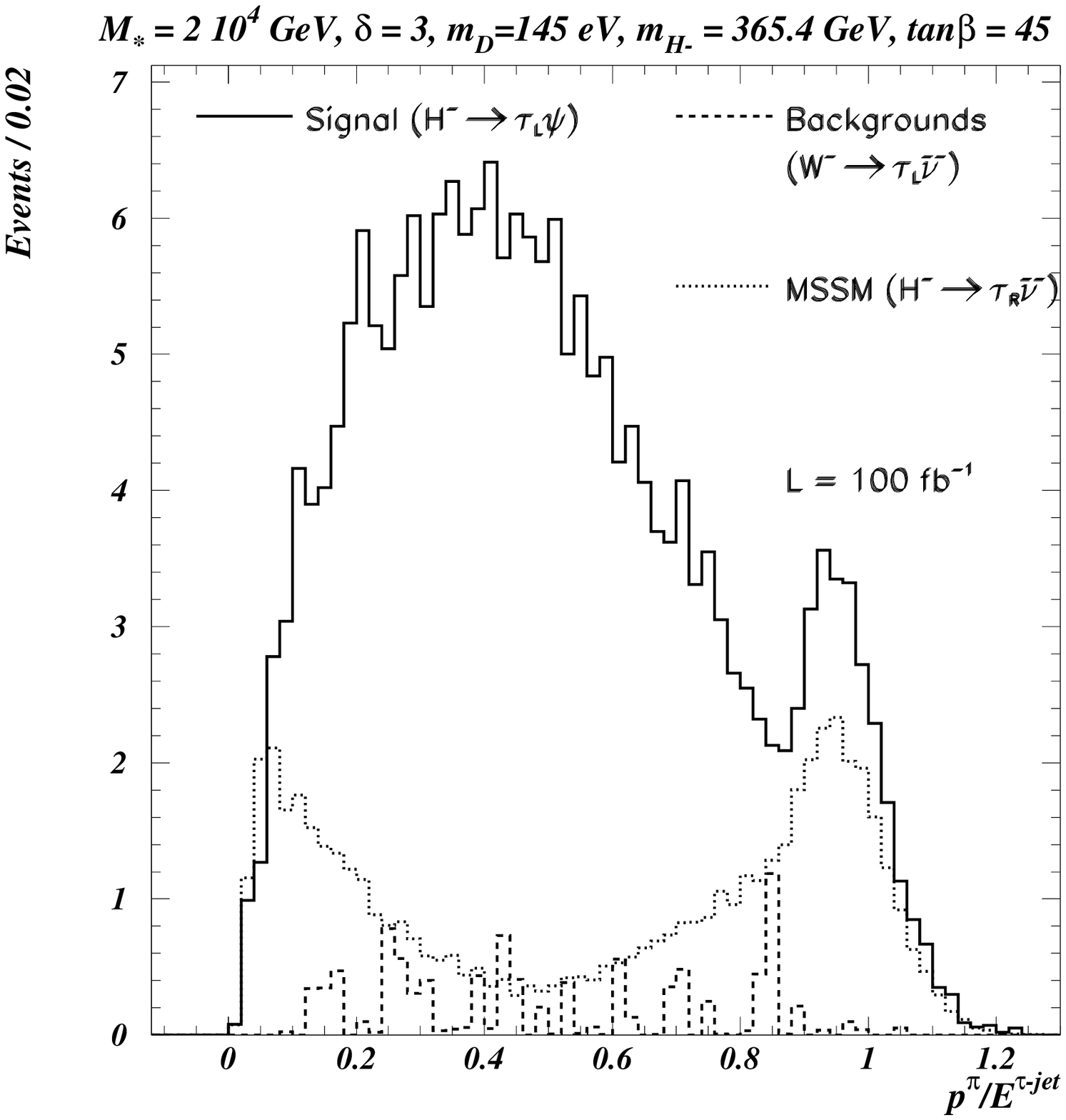} 
\caption{The distribution of the ratio of the charged pion track momentum in one  
prong $\tau$ decay to the $\tau$-jet energy for $m_A=350$~GeV, 
$\tan\beta=45$, $M_*= 20$~TeV, $\delta=3$ and $m_\nu=0.05$~eV. In the 2HDM-II, 
this ratio would peak near 0 and 1 as shown while in other models, the actual 
distribution of this ratio would depend on the polarization asymmetry since 
both left and right handed $\tau$'s would contribute. In the case shown, the 
asymmetry is $\sim 1$ and the ratio peaks near the center of the 
distribution.} \label{fig:350_45_145_1}  
\end{center}  
\end{figure} 
In MSSM, this 
distribution peaks near 0 and 1 while in $H^-\rightarrow\tau_L^-\psi$ from 
large extra dimensions and in the backgrounds, this distribution peaks in the 
center. The backgrounds are relatively very small, and as concluded 
in~\cite{6} and~\cite{7}, the discovery reach is limited by the signal size 
itself. Therefore the observation of a signal in the transverse mass 
distribution and in the distribution of the fraction of the energy carried by 
the charged track should help determine whether the scenario is MSSM or not.   
\end{description}

The main effects responsible for the suppression of the backgrounds are: the 
azimuthal opening angle --- between the $\tau$-jet and the missing transverse 
momentum --- which peaks forward in the backgrounds ($W^\pm\rightarrow\tau\nu$) and backward in the signal ($H^\pm\rightarrow\tau\psi$); and the difference 
in the kinematic bounds on the transverse mass --- this bound is at the W-mass in the backgrounds whereas in the signal, the bound is at the charged Higgs mass. The overall efficiencies of the kinematic cuts (c), (d) and (e) might 
change as a result of the event by event difference in the neutrino mass 
$m_\psi$ leading to an overall change in the signal-to-background ratios and 
signal significances of Table~\ref{tab:table3} but the results of 
Figures~\ref{fig:led_tnu_bgd}, \ref{fig:200_30_0.3ev} 
and~\ref{fig:350_45_145_1} would not be affected because these results rely on 
the differences in the $\tau$-polarization and in the kinematic bounds on the 
transverse mass, irrespective of the neutrino mass $m_\psi$. In Table~\ref{tab:table3}, we present results for 3 different values of the number of extra 
dimensions. With different mass distributions of $m_\psi$ depending on the 
number of extra dimensions, the overall efficiencies for the cuts (c), (d) 
and (e) might change differently for each of the cases presented in 
Table~\ref{tab:table3}.
 
\section{Conclusions} 
 
Large extra dimensions models with TeV scale quantum gravity assume the existence  
of additional dimensions where gravity -- and possibly other fields 
-- propagate. The size of the extra dimensions are constrained to the 
sub-millimeter level since no experimental deviations from the Newtonian 
gravity has been observed at distances larger than $\sim 0.2$ millimeter.  
 
In these models, the right handed neutrino can freely propagate into the extra 
dimensions because it has no quantum numbers to constrain it to the SM brane. 
The interactions between the bulk neutrino and the 
SM fields on the brane can generate Dirac neutrino masses consistent with the 
atmospheric neutrino oscillations without implementing the seesaw mechanism. 
There are no additional Higgs bosons required in these models. The charged 
Higgs productions are therefore the same as in the 2HDM.  
\par 
The charged 
Higgs can decay to both the right and the left handed $\tau$-leptons, 
$H^-\rightarrow\tau_R^-\bar{\nu}+\tau_L^-\psi$ whereas in the 2HDM-II such as 
MSSM, only the right handed $\tau$ decay of the $H^-$ is possible through the 
$\tau$ Yukawa coupling: $H^-\rightarrow\tau_R^-\bar{\nu}$. The $\tau$ decay of 
the charged Higgs has been studied in details for ATLAS and CMS. In the 
current study, we focus on the observability of $H^-\rightarrow\tau_L^-\psi$ 
at the LHC for Higgs masses larger than the top-quark mass. 
\par 
The charged 
Higgs is generated through the $2\rightarrow 2$ process, $gb\rightarrow 
tH^\pm$ --- where $H^-\rightarrow\tau_R^-\bar{\nu}+\tau_L^-\psi$ --- and we 
require the hadronic decay of the associated top-quark: $t\rightarrow jjb$. 
The major backgrounds considered are the single top-quark production, 
$gb\rightarrow tW^\pm$ and the $t\bar{t}$ production with one $W^+\rightarrow 
jj$ and the other $W^-\rightarrow\tau_L^-\bar{\nu}$. We include the $\tau$ 
polarization in the analysis and select one-prong hadronic $\tau$ decays since 
these events carry a better imprint of the $\tau$ polarization. Due to the 
neutrino in the final state, only the transverse mass can be reconstructed. In 
the backgrounds, the transverse mass has an upper bound at the W mass while in 
the signal, the bound is at the charged Higgs mass. As a result, above the 
W threshold, the background is relatively very small. Thus, the discovery 
reach of the charged Higgs in the $\tau\nu$ channel is limited by the signal 
size itself. 
\par
The mass of the neutrino $\psi$ would be different on event by event basis. Consequently, the efficiencies of the kinematic cuts would somewhat be different. However, main results of the current analysis derive from the differences in the polarizations of the $\tau$-lepton and in the transverse mass bounds, and would not be significantly affected by the neutrino mass effect.
\par
Although the observation of a signal in the transverse mass 
distribution can be used to claim discovery of the charged Higgs, it is 
insufficient to pin down the scenario that is realized. Additionally, by 
reconstructing the fraction of the energy carried  by the charged track in the 
one-prong $\tau$ decay, it is possible to claim whether the scenario is the 
ordinary 2HDM or not. The further measurement of the polarization asymmetry 
might provide a distinctive evidence for models with singlet neutrino in large 
extra dimensions. 
 
\begin{acknowledgments} 
K.~A.~Assamagan expresses gratitude to  K.~Agashe for fruitful discussions.  
This work was partially performed at the Les Houches Workshop: ``Physics at 
TeV Colliders'' 21 May -- 1 June 2001. We thank the organizers for the 
invitation. 
\end{acknowledgments}

\end{document}